# Dynamics of H atoms surface recombination in low-temperature plasma

V. Gubarev[1,2,a)], D. Lopayev[3], A. Zotovich[3], M.A. Blauw[4], V. Medvedev[1,2], P. Krainov[1,2], D. Astakhov[2], and S. Zyryanov[3,5]

[1]*Moscow Institute of Physics and Technology, Dolgoprudny, Russian Federation*
[2]*Institute of Spectroscopy of the Russian Academy of Science, Russian Federation*
[3]*Skobeltsyn Institute of Nuclear Physics, Moscow State University, Moscow, Russian Federation*
[4]*ASML, Veldhoven, Netherlands*
[5]*Faculty of Physics, Lomonosov Moscow State University, MSU, Moscow, Russia*

**ABSTRACT**

The dynamics of H atom recombination on materials of interest for a EUV lithographer was studied under a long-term low-pressure $H_2$ plasma exposure. The similarity of the experimental plasma with the typical EUV-induced plasma over the multi-layer mirrors (MLMs) surface of the EUV lithographic machine is demonstrated by means of 2D PIC MC simulation. The measurement of the temporal dynamics of the H atom surface loss probability ($\gamma_H$) is chosen for testing the surface modification during the treatment. Time-resolved actinometry of H atoms with Kr as the actinometer gas was used to detect the dynamics of the H-atom loss probability on the surface of Al, Ru, RVS and $SiO_2$. It is demonstrated that significant changes of the materials surface occur only at the very beginning of the treatment and ist due to the surface heating and cleaning effects. After that no changes of the $\gamma_H$ are found, indicating that the surface stays absolutely stable. A special test of the sensitivity of the used method to the state of surface was carried out. Dynamics of the $\gamma_H$ changes with the small $O_2$ addition clearly demonstrated modification of the Al surface due to oxidation with the next removal of the oxygen by the $H_2$ plasma treatment. The rate of oxide removal is shown to be determined by plasma parameters such as the ion energy and flux to the surface.

## I. INTRODUCTION

Extreme ultraviolet lithography (EUVL) is a crucial stage in the manufacturing of silicon microcircuits with a typical scale of the circuit features below 10 nm [1]. One of the key elements of the EUVL machine is a projection optical system [2,3]. The required lifetime of projection EUV optics should be rather long, up to ~ 30000 hours [4]. The EUV optics is a system of multilayer mirrors, MLMs. One of the main problems limiting the lifetime of EUV optics is contamination by carbon, tin etc. as well as surface oxidation [5,6]. In order to prevent contamination hydrogen gas atmosphere at p ~ 5 Pa is used which provides a cleaning process [1,7,8]. Molecular hydrogen buffer gas reduces tin penetration to the projection optics box. Besides, atomic hydrogen is able to create volatile compounds such as $SnH_4$, $CH_4$, $H_2O$ that can be pumped out. The hot filament sources [9] of atomic hydrogen are commonly used for cleaning. However their application is challenging due to the big thermal loads on multi-layer mirrors and additional chemical pollution by filament material. Moreover, the rate of H cleaning is relatively low and requires large atom doses under the considered conditions.

Another approach for MLMs cleaning is to use EUV-induced plasma itself. During EUV pulse the high-energy photons (hν = 92eV) and secondary electrons, coming from the mirror surface ionize $H_2$ gas creating a plasma [10-12]. Ions and H atoms, coming from the plasma, interact with a lithographer walls and mirrors stimulating surface cleaning [13]. However, the plasma might also have a negative effect causing sputtering and blistering of MLMs as well as modification of EUV chamber materials [14]. Therefore, the knowledge on the behavior of such materials being under a long-time $H_2$ plasma exposure is of a special interest.

As known, the H atom surface loss probability is a sensitive function of the surface state and is able to reflect indirectly complex surface modifications. This work is devoted to the investigation of the dynamics of H atom surface loss probability on several materials of interest for EUV lithography, Al, Ru, RVS and $SiO_2$, under a long-term exposure to low-pressure $H_2$ plasma under the conditions similar to the ones in the EUV-induced plasma.

## II. 2D PIC SIMULATION

Plasma parameters in the EUV-induced plasma can vary in a wide range in dependence on the EUV pulse power density, pulse repetition rate, gas pressure etc. In order to simulate adequately the plasma conditions in the EUVL machine 2D PIC MC model has been developed [15]. As an example, the simulated energy spectrum of ion flux at an MLMs surface for typical EUV-plasma conditions (a normally incident beam with diameter fwhm = 2.4 cm onto a grounded surface, pulse duration 80 ns, pulse energy 1 mJ, $H_2$ pressure 5 Pa) is presented in fig.1. Red, green and blue points denote the ion energy distribution function (IEDF) of $H^+$, $H_2^+$, $H_3^+$ ions respectively; big black points depict the total calculated energy distribution of ion flux (IEDF) at the surface.

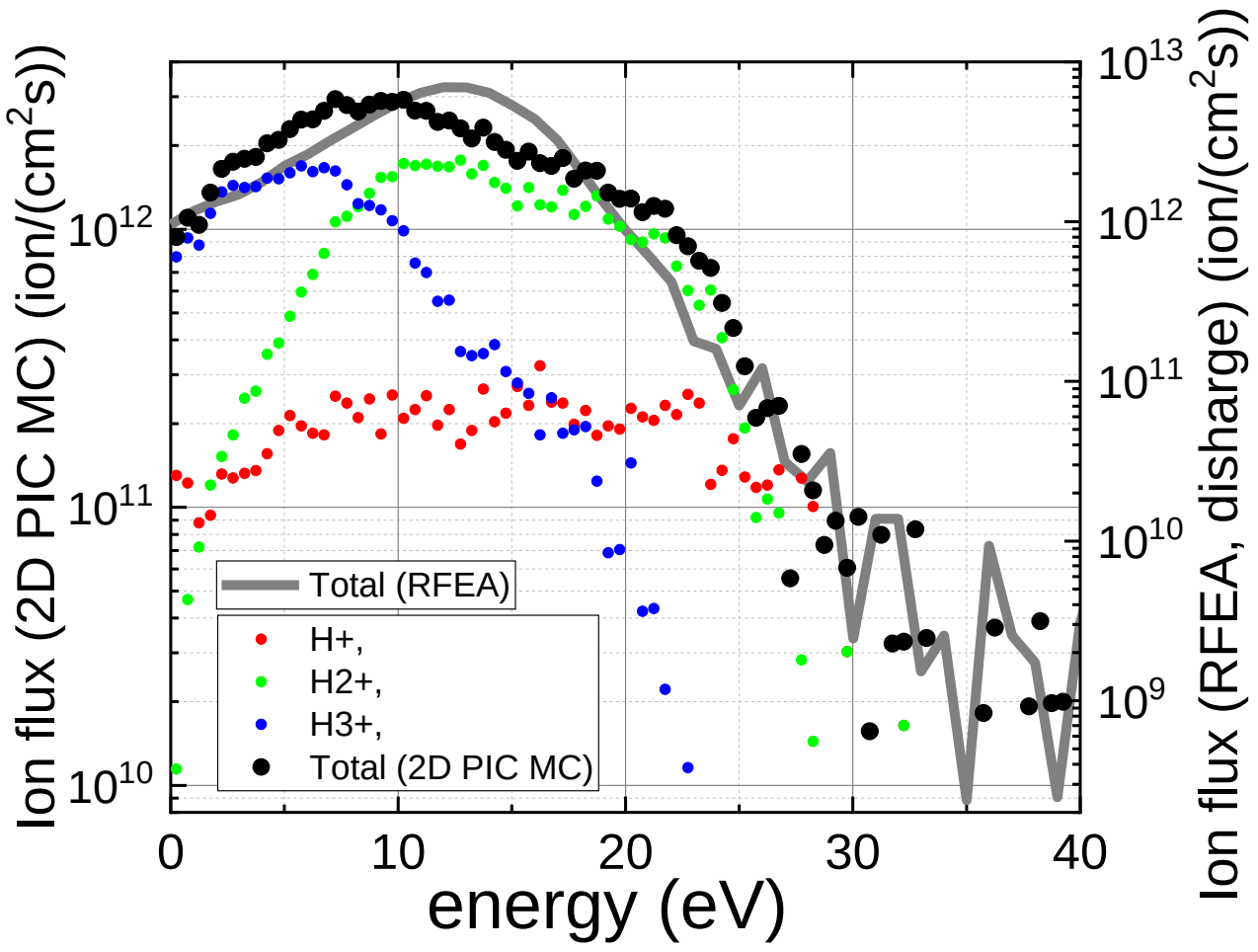

FIG. 1. The ion flux energy distribution (IEDF) (left vertical scale) from simulated EUV-induced plasma and SWD plasma (right vertical scale; see the text). Red blue, green and blue points denote IEDF of $H^+$, $H_2^+$, $H_3^+$ ions respectively, big black points depict the total calculated IEDF. The measured IEDF in SWD plasma is given by grey line. Simulation parameters: $H_2$ pressure is 5 Pa, EUV beam diameter is fwhm = 2.4 cm, EUV pulse energy is 1 mJ, EUV pulse duration is 80 ns. SWD parameters: $H_2$ pressure 40 Pa, input rf power 20W.

Two parts in the simulated spectrum might be distinguished: the fist - closer to 0 eV and the second - above 20 eV. The first one corresponds to the IEDF in the EUV pulse afterglow when plasma is “cold” due to fast cooling of electrons. Such low-energy ions are able to assist the surface reactions with H atoms both coming from a gas phase as well as adsorbed on the surface. The second one corresponds to the IEDF during the EUV pulse when plasma is “hot” due to the continuous creation of a lot of energetic electrons. The total flux of these energetic ions even at a high repletion rate is lower than the flux of cold ions ($F_i < 10^{12}$ 1/(cm$^2$s). However, a large dose of such ions can produce accumulating effect affecting surface properties due to material sputtering and/or modification owing to the hydrogen implantation into the surface layers. Such modification also influences the H atom surface recombination rate. In this work we focus on the evolution of H surface recombination rate at the different materials under a long-term exposure to "hot" H ions to estimate their effect on stability of EUV lithographer operation conditions.

## III. EXPERIMENT

The similarity of the EUV-induced plasma parameters during the pulse with those in an electrode-less low-pressure plasma (like rf ICP and SWD) allows experimental reproducing of the "hot" part of IEDF of the EUV-plasma. In the current study a surface-wave discharge was used to study the dynamics of H atom recombination rate at various materials exposed to $H_2$ SWD plasma. The setup is schematically shown in fig.2. The SWD plasma was created in a quartz tube with a diameter 80 mm and length 50 cm while the metal flanges at the ends of the tube was closed by quartz plates (excepting small slits near the tube wall). The gas pressure varied in a range of 5 - 40 Pa by changing the gas flow rate in the range 5-20 sccm while the leak rate into the chamber was less than 0.005 sccm. Two copper ring electrodes located at distance of 2 cm from each other and driven at 81 MHz rf voltage (with rf power up to 50W) were used for creating the plasma inside the tube. The generator was connected to the electrodes through a matching network. The samples (thin rectangular plates) were placed inside the tube. The time-resoled actinometry of H atoms on Kr was used to measure H atom loss probability. The condenser collected emission from the quasi-neutral plasma volume above the sample and then the emission was focused on the entrance slit of the imaging spectrometer Solar TII MS 3504i (spectral resolution 0.3 nm FWHM) and detected by PMT (Hamamatsu R13456). The signal from PMT was transmitted to the NI DAQ board. The LabView program was used for the automation of the whole experiment: DAQ board measurements, flow meters, rf generator, modulator, database etc.

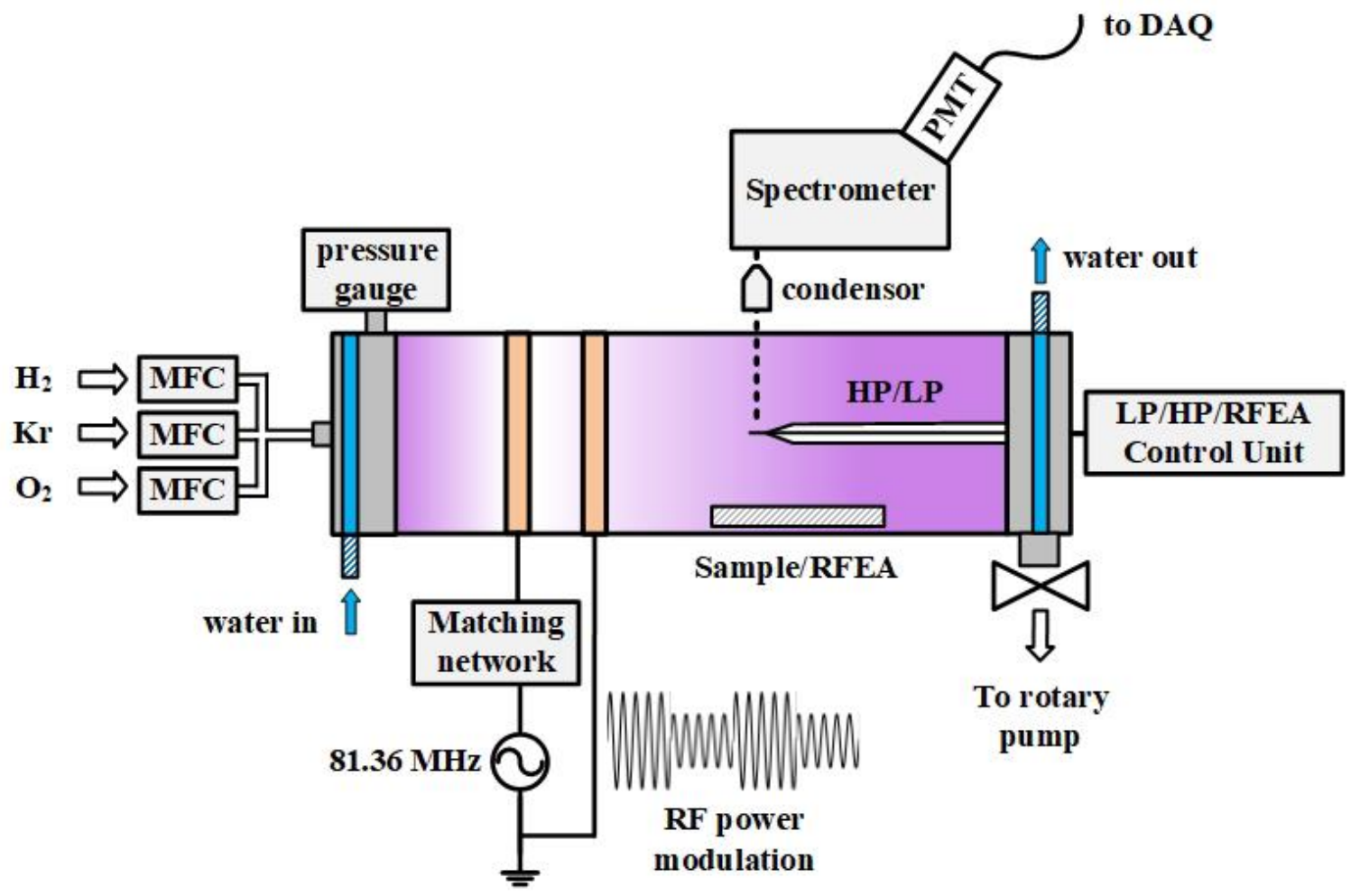

FIG. 2. Scheme of the experimental setup.

The energy spectrum and flux of ions falling on the sample surface were measured with Retarding Field Energy Analyzer (RFEA), Langmuir probe (LP) and MW hairpin probe. A flat rf-compensated RFEA placed at the sample position was used for measuring the ion energy distribution shape, while a Langmuir probe and MW hairpin probes were applied to measure electron temperature and plasma density respectively directly over the sample. The ion flux to the surface was evaluated from plasma density by using the classical formula for the flat collisionless sheath:

$$F_i \approx 0.5 n_e V_B \tag{1}$$

where $V_B = \sqrt{\frac{T_e}{M_i}}$ - Bohm velocity, $T_e$ – electron temperature, $M_i$ - ion mass. We applied the formula (1) for the total pressure range (up to 40Pa) since the plasma sheath for grounded and floating samples is rather thin while the main ion in the considered conditions is $H_3^+$ for which collisional and charge exchange cross sections with $H_2$ are small enough. In a considered range of input rf power, 5-100 W, and $H_2$ pressure, 5-40 Pa, the EEDF body is close to the Maxwellian one. It allowed estimating electron temperature which was varied in a range $T_e$ = 3.6-5.5 eV. Thus, the main variation of ion flux to a sample is provided by varying the plasma density. The plasma density is almost a linear function of input power. The ion flux increases with input rf power and may be varied in the range of $F_i \sim 10^{13}$-$10^{16}$ (ion/(cm$^2$s). It should be noted that some discrepancy was observed in the $n_e$ measured by hairpin and Langmuir probes. The $n_e$ evaluated from Langmuir probe ion saturation current by using OML theory is always higher while from the electron saturation current lower than the $n_e$ measured by the hairpin probe [16]. The correction of the EEDF measured by Langmuir probe by extrapolating the measured EEDF by Maxwellian function to the plasma potential to compensate the electron drain and rf-incompensation effects gives plasma density comparable with the hairpin probe measurements [17] but increases an error. Thus the hairpin probe measurements were used for evaluating ion flux and dose at the sample surface.

The examples of the IEDF measured in SWD plasma for $H_2$ pressure 40 Pa with input rf power 20 W is presented by thick grey solid line in fig.1. It is seen that appropriate choice of discharge conditions allows simulating quite adequately the IEDF shape in EUV-induced plasma. At the same time, the SWD plasma demonstrates higher ion fluxes in comparison with EUV-induced plasma. Thus, SWD plasma conditions allows probing a very long EUV-induced plasma exposures (thousands of hours) of different materials during a reasonable period of time (hours).

Depending on the gas flow rate and input rf power it was possible to vary the IFED peak position from 5 eV up to the 55 eV. The higher pressure, the lower the IEDF peak energy. To treat the materials by high doses of "hot" ions the low-pressure regimes are more desirable due to a much shorter exposition time. Thus, the study was first of all focused on the regimes with gaspressure below 15 Pa.

Four different materials were exposed to $H_2$ SWD plasma. Those are plates of Al and RVS (stainless steel), thin 50 nm Ru films on quartz and $SiO_2$ (quartz chamber itself). The area of the SiO2 sample was 1105cm$^2$; RVS and Al - 50cm$^2$; Ru - 25cm$^2$. For Al, RVS and $SiO_2$ the maximum input power was 50 W, that provided the IEDF peak position at ~50eV with ion flux ~ $3{\cdot}10^{15}$ (ion/(cm$^2$s). In order to avoid overheating and blistering the Ru samples were exposed to weaker plasma at 10W that provided peak position of IED at ~37eV with the total ion flux ~ $1.5{\cdot}10^{15}$ (ion/(cm$^2$s).

**IV. H ATOM LOSS RATE MEASUREMENTS: PULSED ACTINOMETRY METHOD.**

For H atom loss rate measurements the approach proposed in [18,19] was applied. The sharp-edge small modulation of discharge power is used to get information about the slow recovery kinetics of H atom mole fraction under the given conditions, since plasma (both $T_e$ and $n_e$) achieves its steady state during the modulation much faster than H atom density. It allows tracking changes of H density by using actinometry technique.

The concept of actinometry is transparent and well-known [19-21]. A small amount of actinometer (a few percent of a noble gas) is added to the gas flow in order not to disturb plasma parameters. Here 5% Kr was added to the $H_2$ flow. The emission lines of H and Kr are chosen the way that contribution of cascade relaxation and step-wise excitation processes to the corresponding atomic transitions is negligible in the considered conditions. The population and depopulation of the emitting atom states are mostly determined by i) direct excitation by electron impact, ii) dissociative excitation, iii) radiadive decay and iv) collisional quenching. Thus, the ratio between the H and Kr atom densities can be written as follows:

$$\frac{n_H}{n_{Kr}} = \frac{I_{H*}}{I_{Kr*}} C_{Kr}^{H} - \frac{\alpha}{\xi} \qquad (2)$$

where $n_H$ and $n_{Kr}$ are densities of H and Kr atoms in the ground state, $\xi = n_{Kr}/n_{H2}$ - the mole fraction (percent) of krypton, $\alpha = k_{de}/k_e$ – the ratio of the $H^*$ production rate constants by dissociative electron excitation and direct excitations from $H_2$ and H respectively. Under the considered conditions $\alpha$ is quite low that one can neglect the second term in eq.2. Actinometric coefficient $C_{Kr}^{H}$ is defined by equation:

$$C_{Kr}^{H} = \frac{A_{nm} k_e^{Kr} (\Sigma A_{ij} + k_q^{H} n_{H2})}{A_{ij} k_e^{H} (\Sigma A_{nm} + k_q^{Kr} n_{H2})} \qquad (3)$$

where $A_{ij}$ and $A_{nm}$ are Einstein's coefficients for hydrogen and krypton. $k_q^H$ and $k_q^{Kr}$ are rate constants of collisional quenching of the H and Kr emitting levels by molecular hydrogen. $k_e^H$ and $k_e^{Kr}$ are rate constants of electron impact excitation of H and Kr respectively. The excitation rate constants are determined by the excitation cross-sections, $\sigma_e(\varepsilon)$, of the corresponding levels and electron energy distribution function (EEDF), $f(\varepsilon)$:

$$k_e = \left(\frac{2e}{m_e}\right)^{1/2} \int_{E_{ex}}^{\infty} \sigma_e(\varepsilon) f(\varepsilon) \varepsilon d\varepsilon \qquad (4)$$

where $\int f(\varepsilon)\sqrt{\varepsilon}d\varepsilon = 1$, $e$ and $m_e$ are charge and mass of electron. Despite the excitation rate constants itself can noticeably vary depending on the plasma parameters, the ratio $k_e^{Kr}/k_e^H$ remains almost constant with change of plasma parameters if an appropriate choice of H and Kr emitting states are made. If the difference in the excitation energies of the correspondin levels appears to be notably lower $T_e$, then ratio $k_e^{Kr}/k_e^H$ remains almost constant. The following transitions of H atom with wavelength $\lambda$ = 656.6 nm (excitation energy $E_{ex}\approx$12.09 eV) and Kr atom with $\lambda$ = 811.3 nm (excitation energy $E_{ex}\approx$12.06 eV) were used. Thus, knowing the concentration of Kr and measuring the ratio of the chosen H and Kr lines intensities it is possible to evaluate H atoms concentration in the ground state.

Measurements of H atom loss frequency (and thereby H atom surface loss probability in the case when the surface loss is the dominant loss process) can be done by measuring dynamics of H atom concentration during the plasma modulation. To measure the rate of H atom surface recombination under the given conditions the time-resolved actinometry with ~20% plasma modulation has been applied. In the case of the plasma transition from one to another equilibrium states due to slight modulating rf power, the different plasma components reach their equilibrium values with characteristic times individual for each component. The characteristic relaxation times of the electron temperature and plasma density take units and tens of microseconds correspondingly. By contrast, the characteristic time of atomic hydrogen density recovery occurs at a millisecond time scale and determined by the surface loss. The modulation of $T_e$ and $n_e$ can lead to the noticeable modulation of H and Kr atoms emission, while the actinometric signal $I_H/I_{Kr}$ allows ignoring the electron changes after $T_e$ and $n_e$ relaxation time and, therefore, tracking only the dynamics of H density. The dynamics of the actinometry signal was found measuring the dynamics of emission intensities of H and Kr atom lines with subtracting the dynamics of background emission near each line.

In the proposed method it is required to provide the so-called "kinetic regime" when H atom diffusion is fast enough and atom loss is determined by surface loss process. In this case, the H density should have a smoothed spatial profile in the chamber with small concentration gradients.

The hydrogen atom density profile measured by the actinometry method along the reactor tube at pressure p = 15 Pa is shown in fig.3a. The zero position corresponds to the center between the ring electrodes. The region from -3 cm to 3 cm (grey zone) corresponding to the plasma sheath areas (where plasma is not quasi-neutral) is excluded due toincorrect application of actinometric method in this area. H atom density was estimated using eq.2 in assumption of constant gas temperature ~450K. As it is expected, H density profile has a small gradient from the electrodes to the ends of the tube due to the enhanced loss rate at the tube ends because of the metal presence (since H atom loss probability on metal is more than two orders higher in comparis with the one on $SiO_2$). The observed deviation from a constant H density profile is not critical and allows applying the actinometry method for kinetic measurements of H loss rate in the presence of samples in the reactor.

The top part of fig.3b shows an example of intensity dynamics of Kr (green curve) and H (red curve) emission lines and their ratio (blue curve) $I_H/I_{Kr}$ is given in the bottom part. The results shown in fig. 3b correspond to the empty chamber (i.e. for fused silica). The exponential factors for growing and falling parts (red curve in the bottom graph of figure 3b) are almost the same and represent the characteristic loss time of the H atom at the chamber surface. The characteristic diffusion time of H atoms for the considered conditions (15 Pa and gas temperature ~ 450 K) is < 0.3 ms while the measured H loss time ~ 3-4 ms) [22], i.e. the diffusion of H atoms is much faster than the H atom surface loss. So, the loss frequency $\nu_H$ allows evaluating the hydrogen atom loss probability, $\gamma_H$:

$$\nu_H = \frac{S}{V}\frac{\gamma_H \upsilon_{th}^H}{2(2-\gamma_H)} \tag{5}$$

where $S/V$ is the so-called "geometric factor", the ratio of the chamber area to the chamber volume); $\upsilon_{th}^H$ is the thermal velocity of hydrogen atoms; $\nu_H$ is the H loss frequency. The loss frequency for the empty chamber appeared to be rather stable allowing us to measure carefully loss rate changes when samples of Al, Ru or RVS were placed into the chamber. The loss rate at Al, Ru or RVS was then calculated from change in H loss frequency in comparison with the empty chamber with taking into account the area of the exposed materials. To minimize the possible Kr effect (above all, $Kr^+$ ions) on the material surface the Kr was admixed to hydrogen flow only during actinometric measurements. The actiniomeric measurements were fast enough (tens of seconds) that allowed us to study the evolution of the loss probability at different materials during the long-term (hours) plasma exposure.

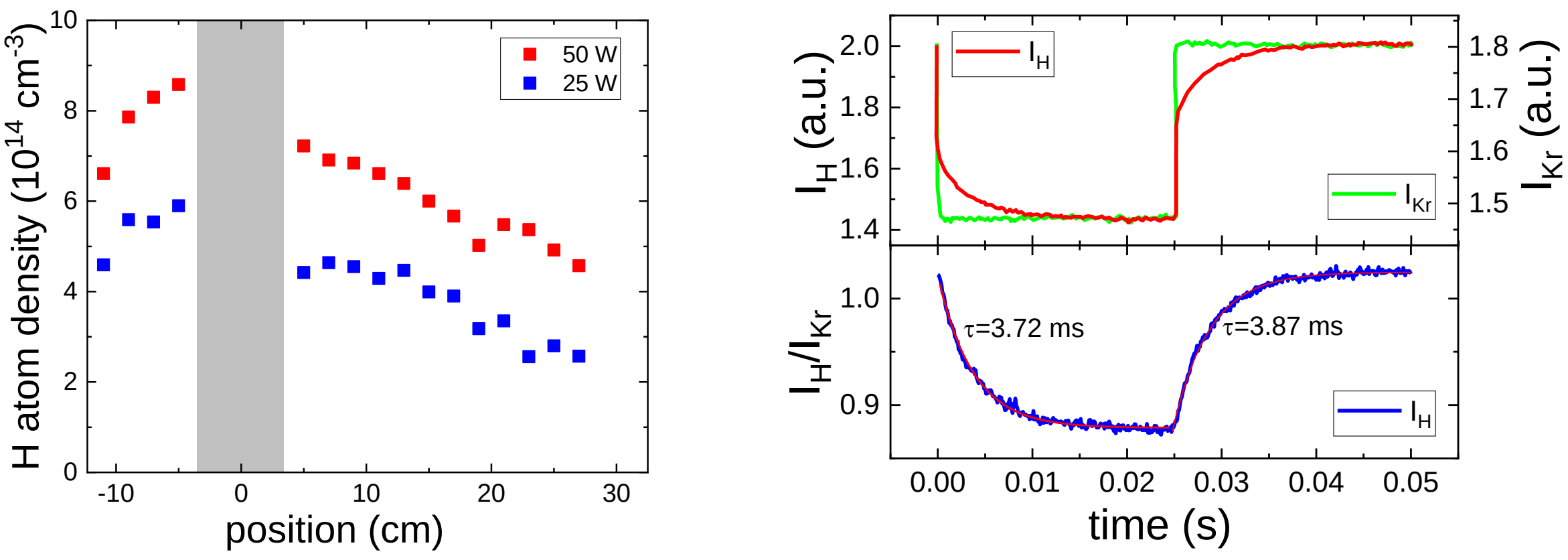

(a) (b)

FIG. 3. a) Atomic hydrogen distribution along the experimental chamber at 15 Pa for SWD powers: 50W and 25W. Grey rectangle shows rf antenna area; b) Top graph - intensity evolution for Kr and H atoms. Bottom one - intensity ratio $I_H/I_{Kr}$. The exponential fitting of increasing/decreasing parts allows measuring the characteristic time of H-atoms surface loss and calculating the hydrogen surface loss probability according to eq.5.

## V. RESULTS AND DISCUSSION

Fig.4a shows the measured evolution of the hydrogen atom surface loss probability ($\gamma_H$) on RVS (blue dots), Ru (green dots), Al (red dots) and $SiO_2$ (black dots) under a long time $H_2$ plasma exposure. The changes in $\gamma_H$ during the first hours can be explained by the surface cleaning effect mainly from contaminations of carbon and oxidized layer. Besides, the samples heating may also promote the increase in H atom loss probability. After the "fluctuations" (during cleaning and heating) the H atom loss probability reaches the plateau for all the materials and remains constant with increasing exposure time. It shows that surface reaches the stable state and $H_2$ plasma doesn't affect it. The values of H atom surface loss probability for the studied materials are presented in Table 1.

Table 1.

| Material | H atom surface loss probability |
|---|---|
| $SiO_2$ | 0.00314 ± 0.00006 |
| Al | 0.089 ± 0.004 |
| Ru | 0.127 ± 0.008 |
| RVS | 0.215 ± 0.005 |

Note here, that the trends for the surface state behavior are quite accurately reproduced by actinometric measurements of $\gamma_H$. However, the obtained absolute values of H atom surface loss probability might have an error up to ~30% because of atom density gradients over the samples and the used approach requires additional validation for obtaining absolute values, which is out of this work scope. Anyway, the obtained results for H atom loss probability on RVS and Al surfaces are in a good agreement with data obtained in the other works, for example in [22]. Thus, fig.4a shows that the long-term exposure to $H_2$ plasma, exactly $H_3^+$ ions with energy below 50 eV, does not lead to a surface modification (except of the cleaning/heating processes) of the studied materials: RVS, Ru, Al and $SiO_2$ which are among the basic materials in EUV lithography tool.

It is also important to emphasize the very high sensitivity of the applied in this work methodology which allows detecting even small changes of surface state of materials. To demonstrate this and the mentioned above surface cleaning the experiment on a long time exposure of Al to plasma with addition of a small amount (5%) of $O_2$ to $H_2$ flow was carried out. As known, oxygen can strongly affect the surface state, first of all, due to oxidation process and hence change the hydrogen atom loss probability. The evolution of H atom loss probability on Al, when oxygen was and wasn't added, is presented in the fig.4b. Vertical dashed lines display the moments of starting/stopping the oxygen flow.

The $\gamma_H$ on Al evolution in fig.4b might be roughly splitted into 8 stages. The discharge power was 50W from 1$^{st}$ to 5$^{th}$ stages. At the 1$^{st}$ stage, the initial stable state of H atom loss probability is observed when oxygen flow is off. At the 2$^{nd}$ stage, the $O_2$ addition leads to the sharp decrease of H atom losses probability (it is seen like a rapid fall in fig.4b) apparently due to adsorption of oxygen atoms which block surface sites for H atom recombination (competitive recombination). At the 3$^{d}$ stage, it replaced with a slower decrease in the H atom loss probability most probably due to Al surface oxidation. The 4$^{th}$ stage corresponds to the moment when oxygen flow was turned off. Now verse visa, a sharp rise of H atom loss probability occurs, apparently due to removal of adsorbed oxygen atoms and, therefore, unblocking sites for H atom recombination. However at the 5$^{th}$ stage the $\gamma_H$ reduction observed at the $3^d$ stage partly continued further probably owing to continuing reconstruction of Al-O surface layer. It goes with gradual slowing down and almost stopping of the $\gamma_H$ change to the $6^{th}$ stage. At the $6^{th}$ stage the $H_2$ discharge power was increased to 75W after that the $\gamma_H$ started to increase due to $H_2$ plasma cleaning of oxidized Al layer (removal of oxygen from the surface layers). At the $7^{th}$ stage, the higher rate of the $\gamma_H$ increase is observed after the input rf power was further increased to 100W. Finally, at the $8^{th}$ stage, when oxygen is fully removed and Al surface is almost recovered, H atom loss probability is stabilized at the level slightly higher than in the beginning at the 1st stage due to the higher temperature of Al sample (because of the higher input rf power - 100W against the 50W in the beginning). Thus the applied method clearly demonstrates high sensitivity to the surface state, for example, to the process of oxygen removal from the oxidized layers by $H_2$ plasma.

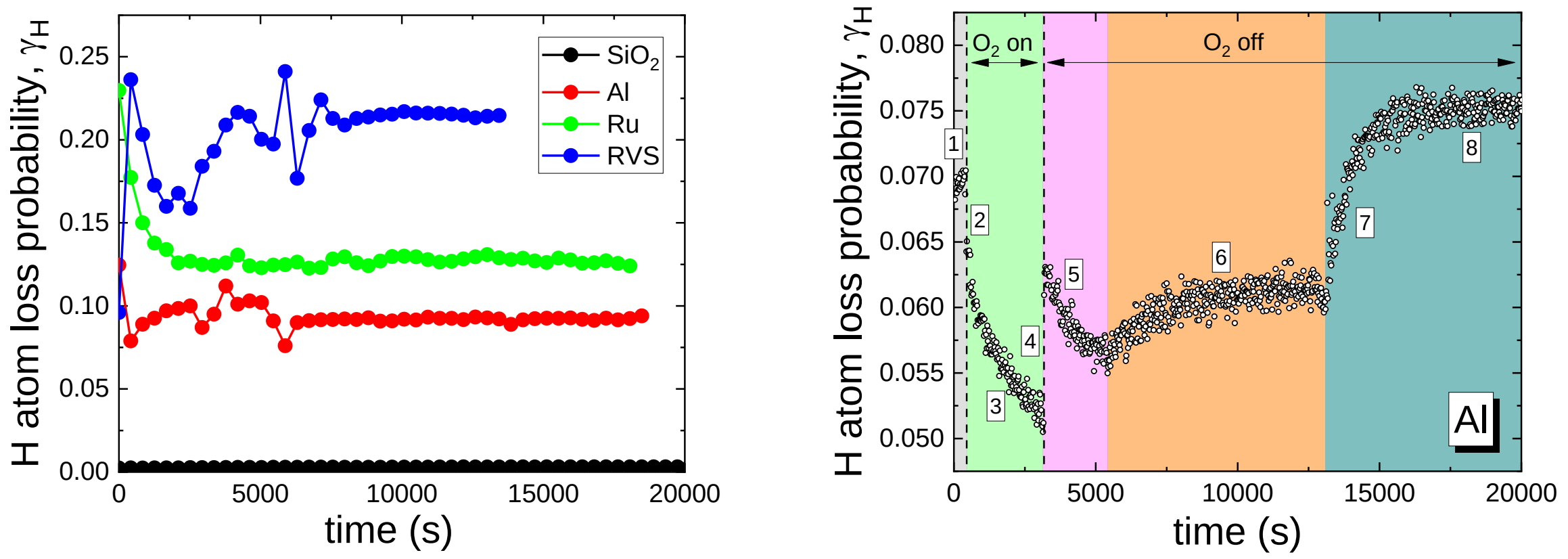

(a) (b)

FIG. 4. a) Evolution of the H atom surface loss probability on RVS (green), Ru (violet), Al (red), $SiO_2$ (blue curve) samples under $H_2$ plasma exposure. The area of the $SiO_2$ sample was 1105cm$^2$; RVS and Al - 50cm$^2$; Ru - 25cm$^2$. b) Evolution of the H atom loss probability on Al surface with addition of 5% $O_2$ (in the time interval t = 280s - 2900s) and variation of the discharge power. Details of the evolution stages are presented in the text. The discharge power is for $1^{st}$-$5^{th}$ stages – 50W, for $6^{th}$ – 75W and for $7^{th}$ and $8^{th}$ – 100W.

## VI. CONCLUSION

In this work, the evolution of the H atom loss probability, $\gamma_{H,}$ on the different materials (RVS, Al, Ru, $SiO_2$) being of interest in EUV lithography was investigated under the long time exposure to low-pressure SWD hydrogen plasma. 2D PIC model was applied to demonstrate that SWD plasma under the applied discharge conditions can adequately simulate the ion flux energy distribution on a mirror surface in EUV-induced plasma. The measurements of $\gamma_H$ clearly show the plasma cleaning effect on all the studied materials during the first two hours. After that the H atom loss probability remains constant under the long time plasma exposure (with ion energy ~50eV and flux ~$10^{15}$ ions/(cm$^2$s) at the sample surface). It shows that after cleaning, $H_2$ plasma doesn't modify the surface of the cleaned materials.

By admixture of a small amount of $O_2$ to $H_2$ it was shown that oxidation decreases the H atom recombination rate. It was also shown that $H_2$ plasma is able to remove oxygen from the oxidized layers and the removal rate is growing with rf power mostly due to increasing of ion energy and flux.

## ACKNOWLEDGMENTS

S. Zyryanov would like to acknowledge Russian Science Foundation: RSF research project No.2172-10040. D. Lopaev, A. Zotovich and S. Zyryanov are also grateful to the Interdisciplinary Scientific and Educational School of Moscow University 'Photonic and Quantum Technologies. Digital Medicine.' for support.